\documentclass[12pt,preprint,aps]{revtex4}

\usepackage{amsfonts}
\usepackage{graphicx}
\usepackage{verbatim} 
\usepackage{multirow}

\begin{document}

\title{Empirical electronic band structure study of 
silver low--index surfaces}
\author{H.J. Herrera--Su\'arez }
\affiliation{
Universidad de Ibagu\'e, 
Facultad de Ciencias Naturales y Matem\'aticas, Colombia,
Carrera 22 Calle 67 Barrio Ambal\'a }
\author{ A. Rubio--Ponce }
\affiliation{
Departamento de Ciencias B\'asicas, 
Universidad Aut\'onoma Metropolitana--Azcapotzalco,
Av. San Pablo 180, M\'exico, D.F. 02200, M\'exico }
\author{ D. Olgu\'in }
\affiliation{
Departamento de F\'isica, 
Centro de Investigaci\'on y de Estudios Avanzados del 
Instituto Polit\'ecnico Nacional,
M\'exico, D.F. 07360, M\'exico}

\begin{abstract}
We studied the electronic band structure of the low--index fcc Ag 
surfaces (001), (110) and (111), by using the empirical tight--binding 
method in addition with the surface Green function matching method. We 
report the energy values for different surface and resonance states 
and compare with the available experimental and theoretical data. 
\end{abstract}

\maketitle

\section{introduction}

In order to predict several surface and bulk crystal properties such 
as mesoscopic equilibrium shape, surface and catalyst activity 
\cite{Stroppa}, growth, creation of rungs and kinks \cite{Baud,Roos}, 
is essential to have a detailed analysis of its electronic band structure. 
Experimental data is complemented with calculation to obtain a deeply 
understand of this systems. Two primary types of calculations are used in 
practice. The first one includes empirical and semi-empirical treatments, 
from wich empirical tight-binding (ETB) is one of the most transparent and 
widely used methods, several recent applications of the ETB method can be 
found in the literature, both for noble metal surfaces \cite{Baud} and elemental 
semiconductors \cite{Velasco}, as more complex systems like the layered II-IV 
compunds \cite{Camara}. The second type are the first-principle calculations 
based on approximate methods such as density functional theory (DFT), these 
are located on the most reliable and widely used methods nowadays.
In this work the electronic band structures of ideal Ag (001), (110) and 
(111) surfaces are discussed. This is a continuation of an extensive study 
of different noble and transition metal surfaces \cite{TesisJHerrera}.

\section{numerical approach}

The ETB calculations were done using a minimal orthogonal basis set. 
Here, a set of nine $``s\ p\ d"$ atomic orbitals
per atom in the unit cell were used, and we have included
the first nearest and next nearest neighbors as proposed by 
Papaconstantopoulos \cite{Papa}.
The parameters of the model are those used by Papaconstantopoulos, it
is known that these parameters properly reproduce the
bulk electronic properties of Ag, according to DFT calculations \cite{Papa}.
To calculate the surface electronic band structure, the
Surface Green Function Matching (SGFM) method was used,
as suggested by Garc\'ia--Moliner and Velasco \cite{GarciaMoliner}. 
The SGFM method, in conjunction with the ETB
approach, was used successfully to study transition
metals \cite{ARubio} and semiconductor surfaces \cite{DOlguin}. 
For a complete formulae of the ETB method
to the formalism of the SGFM 
see details in Refs.~\cite{GarciaMoliner,DOlguin}. 
A recent application of the method to the study of the electronic structure of different 
Pt surfaces has been done. \cite{JHerrera}

From the knowledge of the Green function, the surface and the surface 
resonances states can be calculated from the poles of 
the real part of the corresponding Green function. 
In a similar way, from the imaginary part the local density of states (LDOS) 
can be obtained.

\section{results and discussion}

\subsection{Local density of states}

\begin{figure}[!ht]
\includegraphics[width=0.32\textwidth]{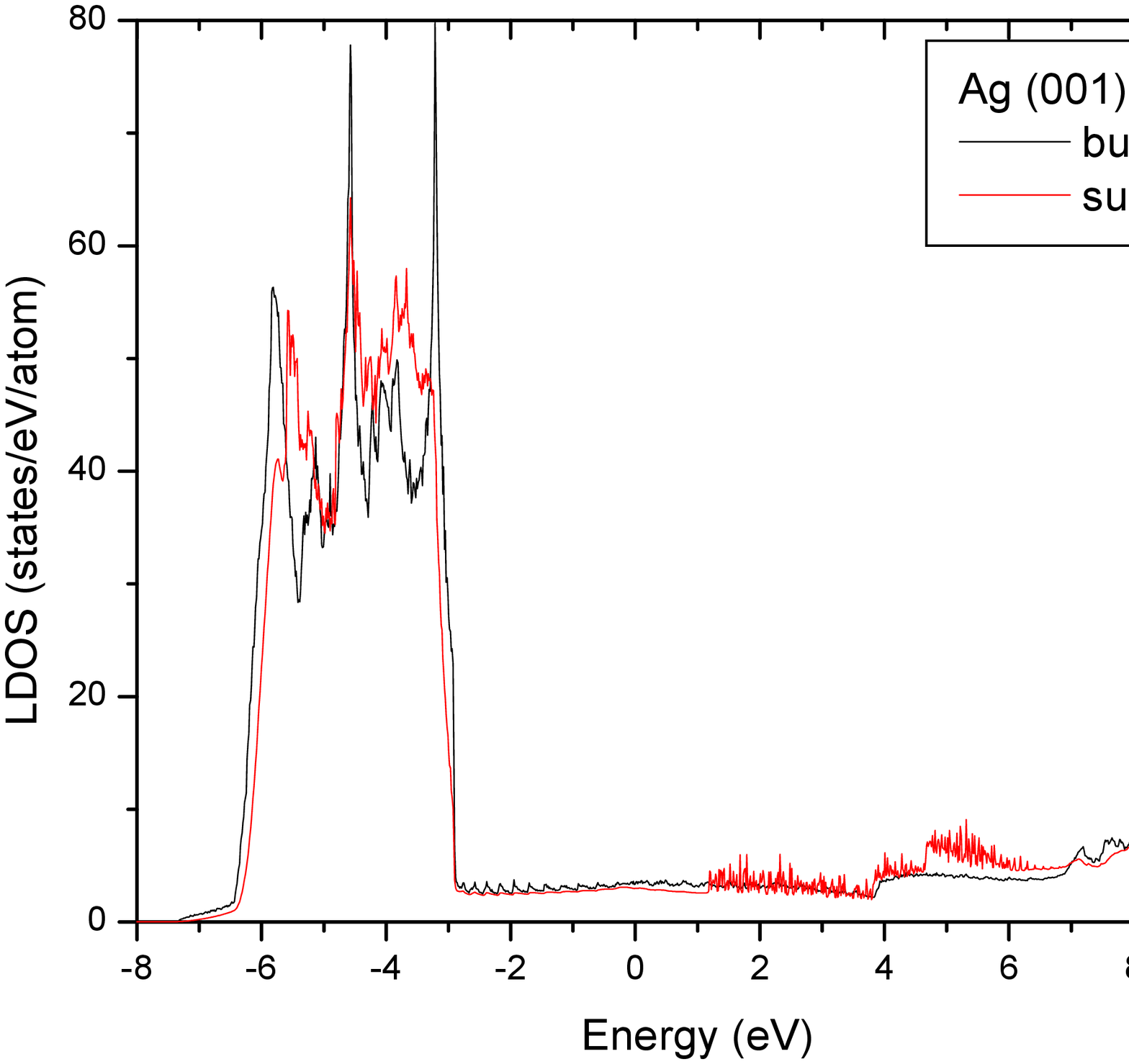}
\includegraphics[width=0.32\textwidth]{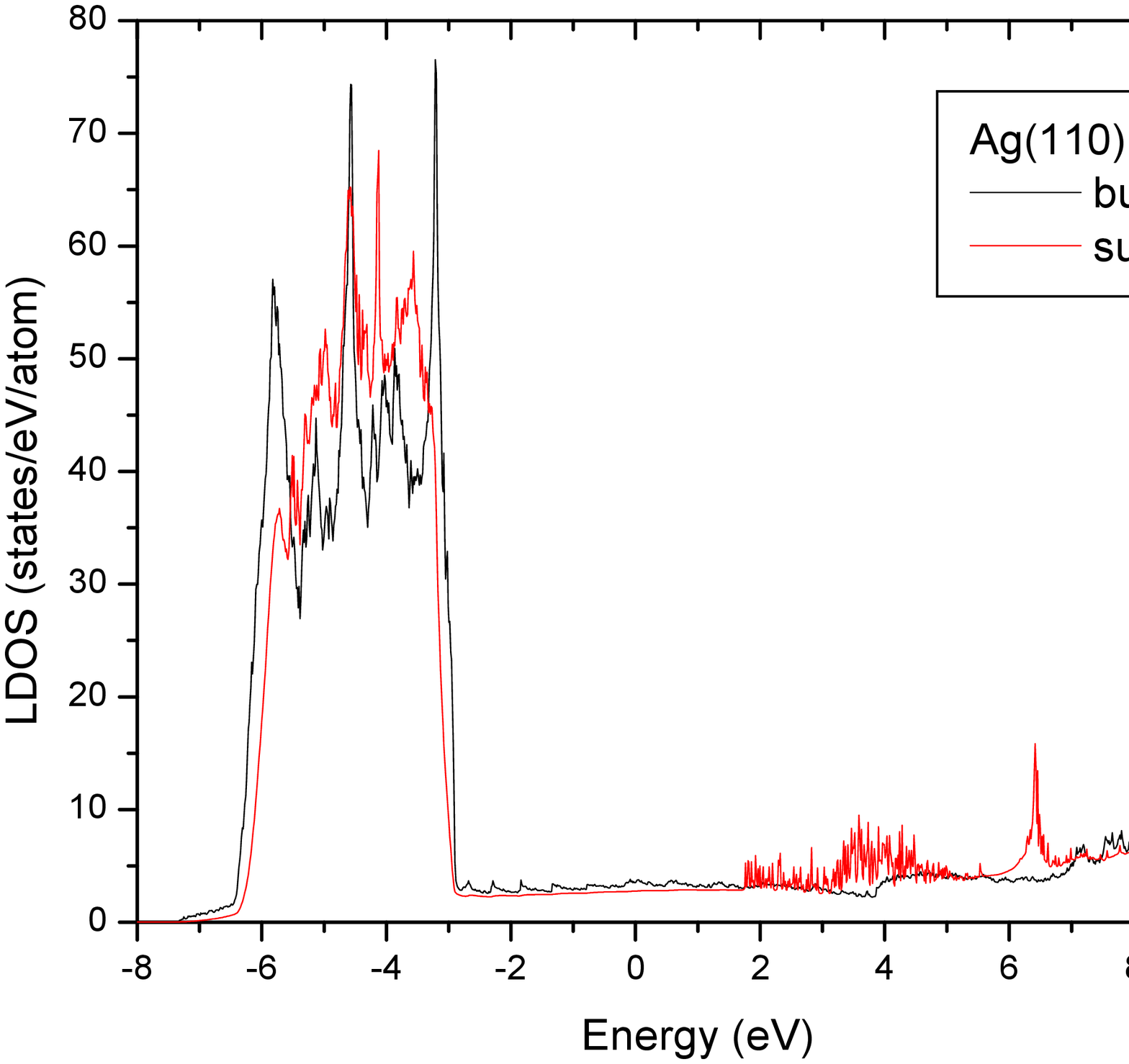}
\includegraphics[width=0.32\textwidth]{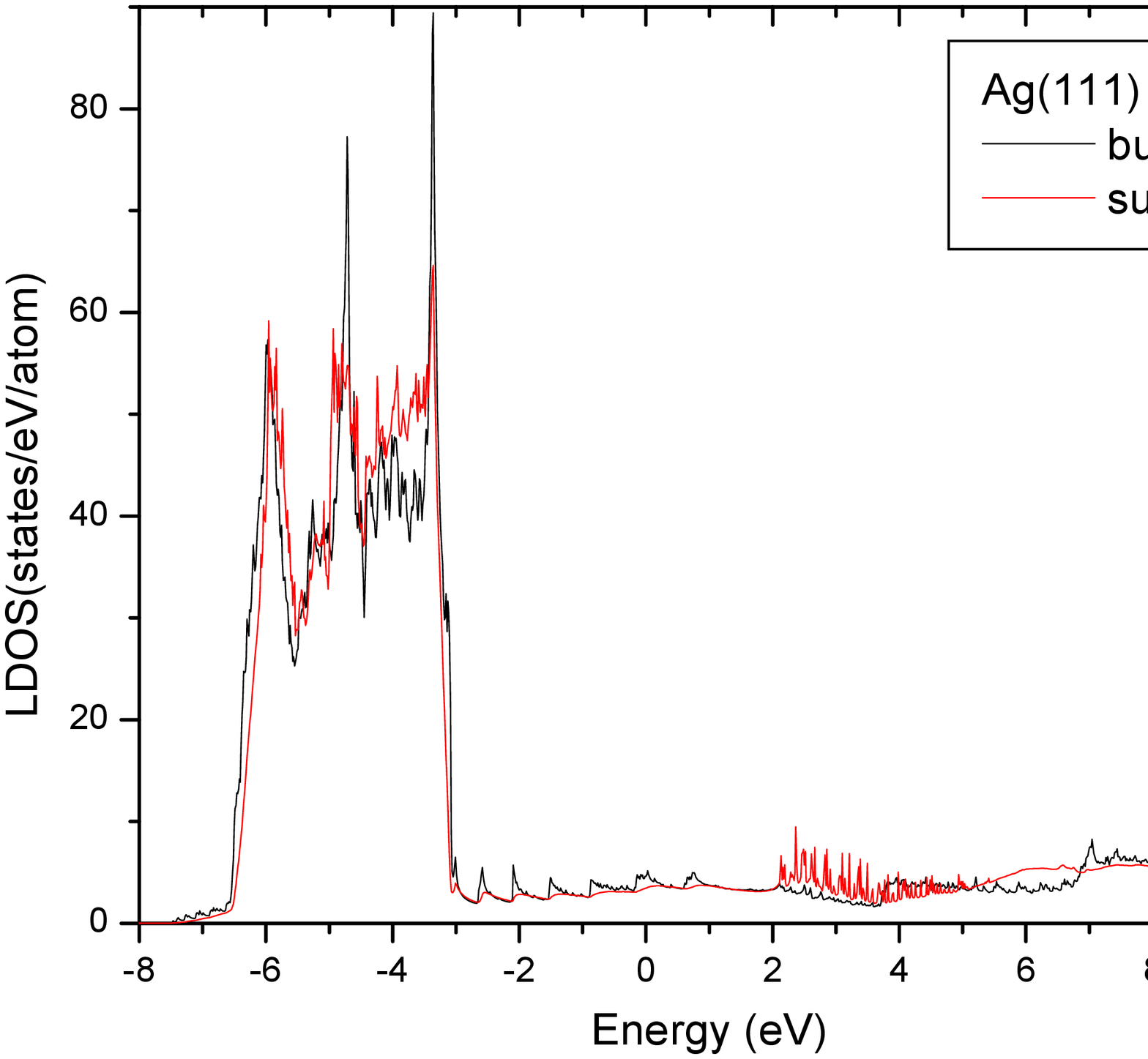}
\caption{(Color online)
Bulk local density of states (LDOS) (black line) 
and surface LDOS (red line) obtained from the SGFM method.
(a) Ag(001)--, (b) Ag(110)--, and (c) Ag(111)--surface.
The zero of energies represents the Fermi level.
} 
\label{Ag001LDOS}
\end{figure}

Figure~\ref{Ag001LDOS} shows our calculated LDOS 
projected on the surface (broken line) and the LDOS projected 
on the bulk (full line), see figure caption for details.

For the Ag(001) surface a total of 528 k-points were used, 
here we used the Cunningham method to found the k-point
set in the irreducible two dimentional segment of the first Brillouin zone
(2D-SBZ) (see inset of Fig.~\ref{Ag001}(a)) \cite{Cunningham}.
For Ag(110) surface we used 256 Cunningham points in
the 2D-SBZ (see inset of Fig.~\ref{Ag001}(b)).
While for the Ag(111) surface we used 136 Cunningham points in the 
2D-SBZ (see inset of Fig.~\ref{Ag001}(c)).

Table~\ref{tablaLDOS} shows the $s,\ p,\ d$  atomic orbital partial 
contribution to the surface and bulk LDOS.
A comparison of our calculations 
with that reported by Papaconstantopoulos \cite{Papa}
shows that our method reproduces properly the bulk DOS.

\begin{table}[ht]
\caption{$spd$ atomic partial contribution to the Local Density
of States (LDOS) at the surface (surf) and bulk (bulk) at the
Fermi Level (states/eV/atom), for the different oriented surfaces
studied in this work} 
\centering 
\begin{tabular}{c c c c c c c} 
\hline\hline 
Surface & $s_{\rm surf}$ & $p_{\rm surf}$ & $d_{\rm surf}$ &
	  $s_{\rm bulk}$ & $p_{\rm bulk}$ & $d_{\rm bulk}$ \\ [0.5ex] 
\hline 
Ag(001) & 0.600 & 0.344 & 9.901 & 0.671 & 0.448 & 9.879 \\ 
Ag(110) & 0.674 & 0.454 & 9.869 & 0.650 & 0.350 & 10.010 \\
Ag(111) & 0.670 & 0.440 & 9.870 & 0.650 & 0.350 & 10.010  \\ [1ex] % [1ex] adds vertical space
\hline 
\end{tabular}
\label{tablaLDOS} 
\end{table}

\subsection{\rm Ag (001)}

\subsubsection{Projected bulk electronic band strucure, 
surface-- and resonance--states}

In an early experimental work Kolb {\it et al.} \cite{Kolb81},
by using electroreflectance in the infrared frequency range, 
reported two surface states (SS) for Ag(001) located in the $\bar X$ high 
symmetry point of the two dimensional Brillouin zone:
the first SS was reported at 3.1 eV above the Fermi level,
while the second state was reported a few meV below the Fermi
level (see Fig. 2 on Ref.~\cite{Kolb81}).
In the same work 
it was found that both states were in good agreement with 
{\it ab initio} pseudopotential calculations.

Then, Altmann {\it et al.} \cite{Altmann} by using angle-resolved 
Bremsstrahlung isochromat spectroscopy corroborate the states
found by Kolb {\it et al.},\cite{Kolb81} and reported other
high energy surface states. 

More recently, Savio {\it et al.}\cite{LSavio} by using 
an {\it ab initio} pseudopontential calculations combined with 
ultraviolet photoemission spectroscopy technique refined the 
previous experimental values for the surface states.

\begin{figure}[!ht]
\includegraphics[width=0.30\textwidth]{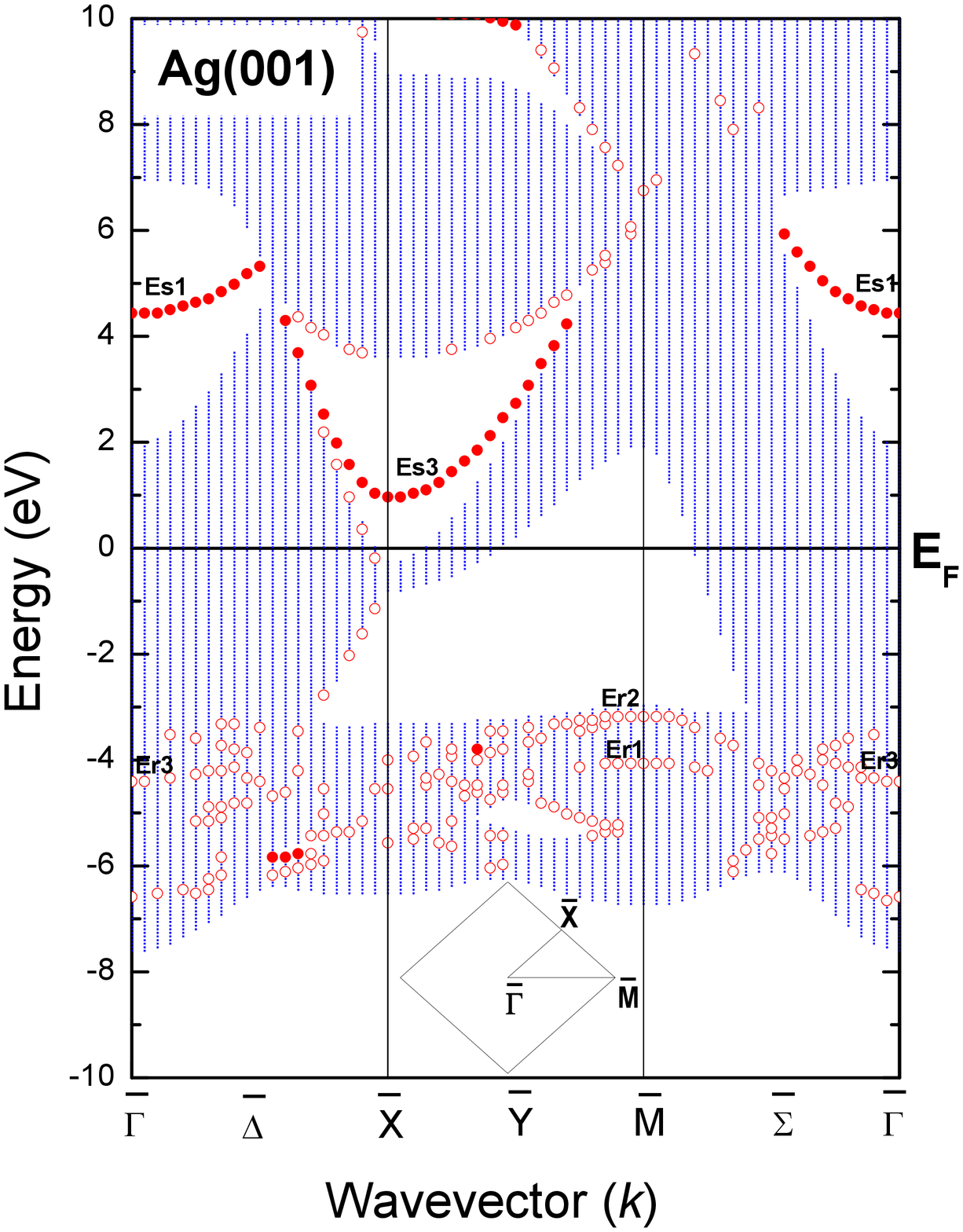}
\includegraphics[width=0.30\textwidth]{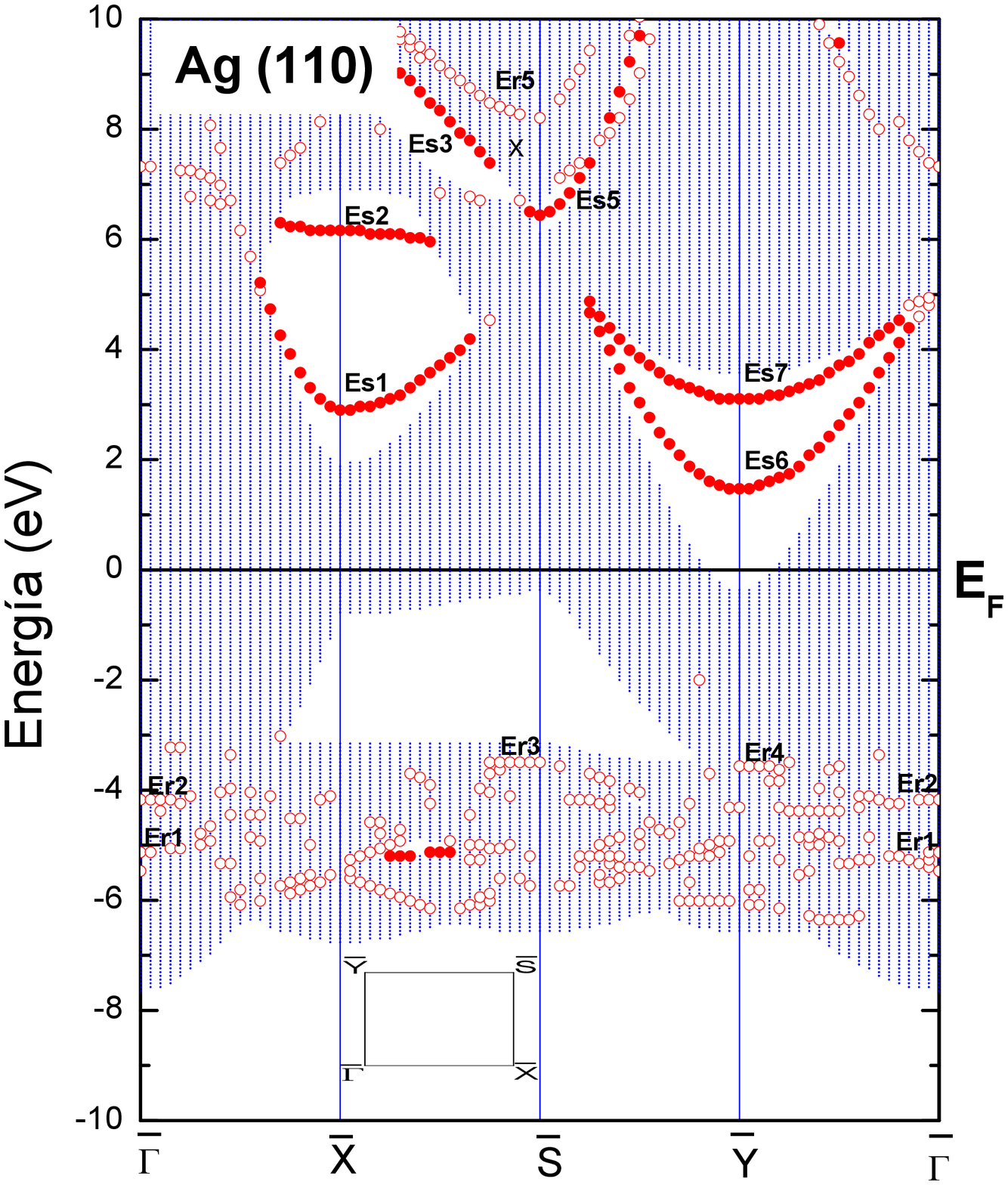}
\includegraphics[width=0.30\textwidth]{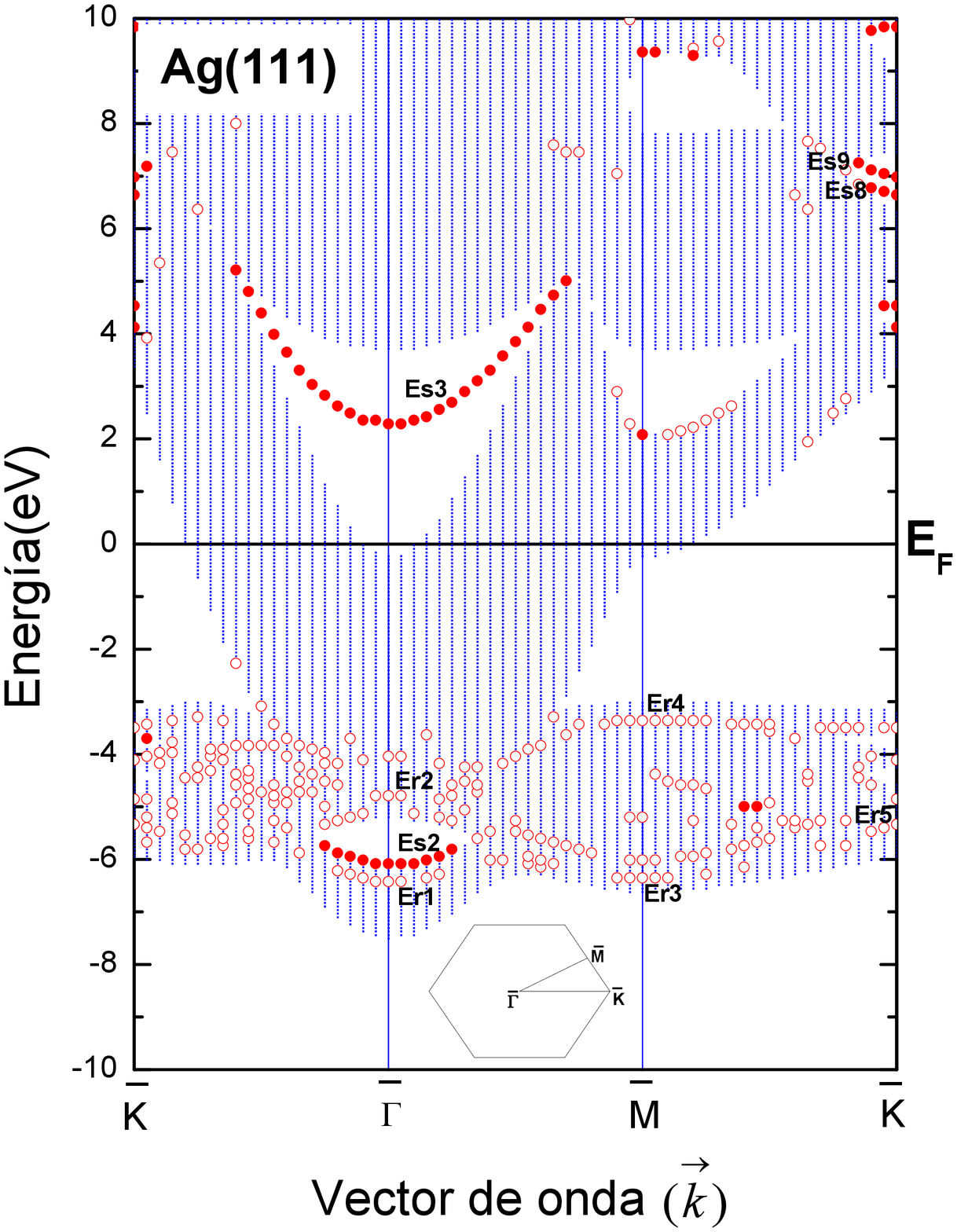}
\caption{(Color on line)
Projected bulk band structure found in our calculation (blue zone).
The full (red) points represent the SS, while the empty points are for 
the RS. 
(a) Ag(001)--, (b) Ag(110)--, and (c) Ag(111)--surface.
The Fermi level is the zero of the energies.
}
\label{Ag001} 
\end{figure}

Figure~\ref{Ag001} shows our calculated
projected bulk band structure and the 
surface-- and resonance--states (RS)
obtained from the poles of the real part of the bulk Green function and 
from the surface Green function, respectively.
As we found, there are two SS labeled 
Es$_{1}$ and Es$_{3}$ (full points), 
and three resonance states
labeled Er$_{1}$, Er$_{2}$, and Er$_{3}$ (empty points). 
The shadow zones represent the calculated bulk band structure projected
on the (001) surface.
We found that our calculated projected bulk bands are in good 
agreement with that reported in Refs. \cite{Kolb81,LSavio}.
Tables~\ref{tablaSS001} and \ref{tablaRS001} list our 
calculated energies for the SS and RS 
found in this work and compare them with other reports.
The states $E_{S5}$ and $S_{S6}$ reported
in the literature, 
are listed in Table~\ref{tablaSS001} 
although our calculations do not reproduce 
these states.

\subsubsection{Detailed discussion of the found SS and RS}

\begin{table}[!ht]
\caption{Surface states for Ag(001).
The first column is the found SS, the second column list
the $k$-point where the SS was found,
the third column shows the 
experimental energy reported for the related state,
the fourth list the theoretical energy value reported
in the literature for the SS, the next column shows our calculated
energy value for the found SS, finally the last column shows 
the symmetry of the atomic orbitals that form the SS, according
to our calculation.
The $k-$vector is given in units of $\left[ \frac{\pi}{a}\right]$,
while the energies are in eV.
}
\centering 
\begin{tabular}{c c c c c c} 
\hline\hline 
SS & $\vec k$ & E$_{\rm exp}$ & E$_{\rm theo}$ &
	  E$_{\rm our}$ & SFO$_{\rm our}$ \\ [0.5ex] 
\hline 
Es$_1$ & (0,0) & 4.00\cite{Altmann} &  & 4.43 & $s,p_z$ \\ 
%Es$_2$ & ($\frac{27}{20},\frac{13}{20}$) &  &  & --3.79 & $d$ \\
Es$_3$ & ($1,1$) & 0.07\cite{LSavio} & -0.13\cite{Kolb81} & 0.96 & $p_x,p_y$ \\
       &         &                   & -0.15\cite{LSavio} &      & \\
%Es$_4$ & ($\frac{11}{20},\frac{11}{20}$) &  &  & $-5.83$ & $s,d_{xy}$ \\
 &  & 3.80\cite{Reihl} &  &  &  \\
Es$_5$ & (1,1) & 3.50\cite{Altmann} & 3.50\cite{Kolb81} &  &  \\
       &       & 3.03\cite{LSavio} & 2.99\cite{LSavio}&    & \\
Es$_6$ & (1,1) & $-0.53$\cite{LSavio} & $-0.53$\cite{LSavio} &  &  \\ [1ex] % [1ex] adds vertical space
\hline 
\end{tabular}
\label{tablaSS001} 
\end{table}

\begin{table}[!ht]
\caption{Resonance states for the Ag(001) surface.
The first column shows the labeled resonance state, the 
second one shows the wavevector of the state in units of 
$[\frac{\pi}{a}]$, the next column shows the related energy in eV, 
finally the last column shows the wave symmetry found 
for the different states.
} 
\centering 
\begin{tabular}{c c c c} 
\hline\hline 
E$_r$ & $\vec k$ & E$_{\rm our}$ & SFO$_{\rm our}$ \\ [0.5ex] 
\hline 
Er$_1$ & (2,0) & --4.06 & $d_{3z^2-r^2}$ \\
Er$_2$ & (2,0) & --3.18 & $d_{xy}$ \\
Er$_3$ & (0,0) & --4.40 & $d_{3z^2-r^2}$  \\ [1ex] % [1ex] adds vertical space
\hline 
\end{tabular}
\label{tablaRS001} 
\end{table}

As we have found the SS Es$_{1}$ 
is located at 4.43 eV at $\bar \Gamma$ showing a parabolic 
dispersion. 
The state is located in the $\bar\Gamma-$gap which has a width of 
5.03 eV, approximately, 
according to our calculations this state is the hibridization
of the $s,p_z$ atomic orbitals.
This SS was measured at 4.0 eV by Altmann {\it et al.} \cite{Altmann}

We found that the SS $E_{S3}$ is located at 0.96 eV at $\bar X$ 
showing a parabolic dispersion,
the state is located in the $\bar X-$gap which width is 3.87 eV.
We found that this state has the symmetry
of the $p_x,\ p_y$ wavefunctions.
The state was theoretically reported by Kolb {\it et al.} \cite{Kolb81},
and was recently mesured and calculated by 
Savio {\it et al.}\cite{LSavio}

The RS Er$_{1}$ was found at $-4.06$ eV in the 
$\left| \text{\={X}\={M}}\right| $
direction and shows almost zero dispersion. 
We conclude that the state is the hibridization of the 
$d_{3z^{2}-r^{2}}$ atomic orbitals.

The RS Er$_{2}$ begins at $-3.18$ eV in the $\bar M$ point 
showing a small dispersion,
as we have found that the character of this state is $d_{xy}$.

The RS Er$_{3}$ is located at --4.40 eV at $\bar \Gamma$,
the symmetry found for this RS is
$d_{3z^{2}-r^{2}}$.

The SS's found in this work for Ag(001) Es$_{1}$ and Es$_{3}$ 
were reported previously in the literature, and we have found 
that our calculated energy values for these states 
show good agreement with published values, as we show
in Table~\ref{tablaSS001}.
While the new states found in this work are the resonant states
$E_{r1}$, $E_{r2}$, and $E_{r3}$.
On the other hand, we do not found the SS E$_{S5}$ and 
E$_{S6}$ listed in Table~\ref{tablaSS001} and reported in the 
Refs.~\cite{LSavio,Reihl,Altmann}.

\subsection{\rm Ag(110)}

\subsubsection{Projected bulk band structure, surface-- 
and resonance--states}

In an early {\it ab initio} calculation Ho {\it et al.}\cite{Kai}
reported the projected bulk band structure and SS of Ag(110),
and predicted a SS on the upper band 
energy region of the 2D BZ
(see Fig.~1(b) in Ref.~\cite{Kai}).
After that, Reihl {\it et al.} \cite{Schlitter}
using the experimental technique of $k-$resolved inverse photoemission
spectroscopy found an unoccupied SS at the energy of 1.65 eV
for Ag(110), that matches in good agreement with the state predicted 
by Ho {\it et al.}, in the energy gap around the $\bar X$ point of 
the 2D-BZ (see Fig.~2 in Ref.\cite{Schlitter}).

Figure~\ref{Ag001}(b) shows our calculated projected bulk band structure 
for the Ag(110) surface, wich shows eight SS (full points) and six RS 
(empty points).
The details of the calculated surface-- and resonance--states
are showed in Tables \ref{tablaSS110} and \ref{tablaRS110}.

\subsubsection{Detailed discussion of the found SS and RS}

As we found the state Es$_{1}$,  
that appears in the upper energy gap located at $\bar X$,
begins at 5.21 eV and ends at 4.19 eV in the interval
$\bar \Gamma- \bar X- \bar S$, crossing the $\bar X$ point at
2.89 eV, 
according to our calculations the state has $p_x$ symmetry of the atomic orbitals.
Although there is not experimental evidence for this SS,
from a theoretical point of view this state has been reported 
in the Refs. \cite{Kai, Kolb81, Tjeng}.
We found that the calculated energy for this state is within the 
reported values (see Table \ref{tablaSS110}).

Moreover, we found another SS in the same gap energy (E$_{S2}$). 
The state shows few dispersion, and it is located around 6.2 eV
on the $\bar X$ point.
Our calculated wavefunction for this state has the 
$s,\ p_z$ symmetry of the atomic orbitals.
From the experimental point of view this SS was reported 
by Altmann {\it et al.} \cite{Altmann} and was found around 
5.0 $\pm$0.2 eV, while the one calculated by Ho {\it et al.},
\cite{Kai} was located at 4.25 eV .

Also, we found that the state Es$_{3}$ shows a great slope, and 
it is located in the $\bar X-\bar S$ interval at energies 
that range from 8.0--10.0 eV.
The wavefunction for this state has the $p_y-p_x$ symmetry of the 
atomic orbitals. 
We do not find experimental evidence for this SS, although the 
state was reported in a theoretical work by Tjeng {\it et al.}
\cite{Tjeng}

The SS Es$_{5}$ is located at 6.05 eV in the $\bar S$ point and shows a great slope, 
the state was found in a energy gap of approximately 
0.54 eV. 
As we found the state is an hybridization of the 
$s,\ p_z$ atomic orbitals.

In the energy gap at $\bar Y$ point, we found two SS 
showing a parabolic shape.
The lower state (E$_{S6}$) is located at 1.46 eV.
Our wavefunction for the E$_{S6}$ state has the $p_y$ symmetry.
This SS was found experimentally in Refs.\cite{Altmann,Bartynski,Tjeng},
and theoretically in Refs. \cite{Kai,Kolb81}.
The upper state (E$_{S7}$) is located at 3.1 eV, and the state has the symmetries 
$s,\ p_z$.
Experimentally this state was reported in Refs.\cite{Altmann,Schlitter,Tjeng},
and theoretically in Ref.\cite{Kai}.
The energy difference found betwen our calculation and the reported
values can bee seen in Table~\ref{tablaSS110}.

\begin{table}[ht]
\caption{Calculated energy values for SS on the Ag(110) surface. 
The first column is the found SS, the second column list the $k$-point 
where the SS was found, the third column shows the experimental 
energy reported for the related state, the fourth list the theoretical 
energy value reported in the literature for the SS, the next column 
shows our calculated energy value for the found SS, finally the last 
column shows the symmetry of atomic orbitals that form the SS, 
according to our calculation. The $k-$vector are given in units of 
$\left[ \frac{\pi}{a}\right]$, while the energies are in eV.}
\centering 
\begin{tabular}{c c c c c c} 
\hline\hline 
surface state & $\vec k$ & E$_{\rm exp}$ & E$_{\rm theo}$ &
	  E$_{\rm our}$ & SFO$_{\rm our}$ \\ [0.5ex] 
\hline 
Es$_1$ & ($\sqrt{2}$,0) & -- & 4.18\cite{Kai} & -- & $-$ \\ 
       &                &    & 4.21\cite{Kolb81} & 2.89 & $p_x$ \\
       &                &    & 2.04\cite{Tjeng} &  &  \\
Es$_2$ & ($\sqrt{2},0$) & 4.86\cite{Altmann} & 5.26\cite{Tjeng} & 6.16 & $s,p_z$ \\
Es$_3$ & ($\sqrt{2},0.4$) & -- & 8.06\cite{Tjeng} & 9.22 & $p_{y,z}$ \\
%Es$_4$ & -- & -- & -- & $-5.19$ & $d_{xy,x^2-y^2,3z^2-r^2}$ \\
Es$_5$ & ($\sqrt(2)$,1) & -- & -- & 6.43 & $s,p_z$ \\
       &       & 0.0\cite{Altmann} & 1.25\cite{Kai} &       \\
ES$_6$ & (0,1) & --0.10\cite{Bartynski} & 1.15\cite{Kolb81}& 1.46  & $p_y$ \\
       &       & 0.10\cite{Tjeng} &  & -- & -- \\ [1ex] % [1ex] adds vertical space
       &  & 1.50\cite{Altmann} & 1.28\cite{Kai} & 3.10 & $s,p_z$ \\
Es$_7$ & (0,1) & 1.65\cite{Schlitter} &             & &    \\
       &       & 1.95\cite{Tjeng} &             & &    \\
%Es$_8$ & $(\frac{\sqrt{2}}{2},1)$ &  &  & 3.03 & $p_y$ \\
%Es$_9$ & $(\frac{\sqrt{2}}{2},1)$ &  &  & 3.85 & $s,p_{x,z} $ \\
\hline 
\hline 
\end{tabular}
\label{tablaSS110} 
\end{table}

\begin{table}[ht!]
\caption{
Calculated energy values for the RS of the Ag(110) surface. 
The first column shows the labeled resonance state, the 
second one shows the wavevector of the state in units of 
$[\frac{\pi}{a}]$, the next column shows the related energy in eV, 
finally the last column shows the wave symmetry found 
for the different states.}
\centering 
\begin{tabular}{c c c c c} 
\hline\hline 
E$_r$ & $\vec k$ & E$_{\rm theo}$ & E$_{\rm our}$ & SFO$_{\rm our}$ \\ [0.5ex] 
\hline 
Er$_1$ & (0,0) & --4.74\cite{Kai} & -5.12 & $d$ \\
Er$_2$ & (0,0) & --3.76\cite{Kai} & -4.17 & $d$ \\
Er$_3$ & ($\sqrt{2}$,0) & --3.48\cite{Kai} & -3.40 & $d_{x^2-y^2}$ \\ 
Er$_4$ & (0,1) & --3.48\cite{Kai} & -3.50 & $d_{xy,zx}$ \\
Er$_5$ & ($\sqrt{2}$,0) & 7.52\cite{Tjeng} & 8.20 & $p_{y}$ \\
Er$_6$ & ($\frac{\sqrt{2}}{2}$,0) & 6.77\cite{Tjeng} & 6.16 & $s,p_{x}$ \\
Er$_7$ & (0,0) &  & -5.46 & $d_{yz,x^2-y^2}$ \\
Er$_8$ & (0,0) &  & 7.31 & $p_z$ 
\\ [1ex] % [1ex] adds vertical space
\hline 
\end{tabular}
\label{tablaRS110} 
\end{table}

In the VB region we found four RS, and according to our calculations
the states are: 1) The RS Er$_{1}$ is located at $\bar \Gamma$ at -5.12 eV,
and its wavefunction has the full symmetry of the $d$ atomic orbitals.
From the theoretical point of view, the state was reported in Ref.~\cite{Kai}. 
2) The RS Er$_{2}$ is located at $\bar \Gamma$ at --4.17 eV, and its wavefunction
has also the full symmetry of the $d$ atomic orbitals.
The state was calculated in Ref.~\cite{Kai}.
3) The RS Er$_{3}$ is located at -3.40 at the $\bar S$ point,
and its wavefunction has the symmetry of the $d_{x^{2}-y^{2}}$
atomic orbitals.
Theoretically the state was reported in Ref.~\cite{Kai}.
4) The RS Er$_{4}$ is found at -3.50 eV at $\bar Y$ point, and its 
wavefunction is an hybridization of the $d_{xy},\ d_{zx}$ atomic
orbitals.

Then, in the CB we found the RS Er$_{5}$,
the state is located at 8.20 eV in the $\bar S$ point, and 
it wavefunction has the $p_y$ symmetry.
The state was calculated in Ref.~\cite{Kai}.

The RS Er$_{6}$ is located at $\bar \Gamma$ at -5.46,
and its wavefunction has the hybridization of the $s,\ p_x$
atomic orbitals.
This state was also calculated in Ref.~\cite{Kai}.

The RS Er$_{8}$ is located at $\bar \Gamma$ at 7.31 eV,
and has the symmetry of the $p_z$ atomic orbitals.

As we found our calculated SS and RS for Ag(110) Es$_{1},$ Es$%
_{2},$ Es$_{3},$ Es$_{6},$ Es$_{7},$ Er$_{1},$ Er$_{2},$ ..., Er$_{6}$, 
have been reported previously in the literature, and in the usual 
precision of the method, show good agreement with them, as we can see in 
Tables~\ref{tablaSS110} and \ref{tablaRS110}.
The new states found in this work are Es$%
_{5},$ Es$_{8},$ Es$_{9},$ Es$_{10},$ Es$_{10},\ $Er$_{7}$ y Er$_{8}.$

\subsection{\rm Ag(111)}

\subsubsection{Projected bulk band structure, surface states,
and resonance states}

In an early calculation Ho {\it et al.} \cite{Ho1983} 
by using the {\it ab initio} pseudopotential method, reported 
the bulk band structure, SS and RS for Ag(111).
They predicted a SS just above the Fermi level in the $\bar \Gamma$
point.
Then in a photoemission measurement Kevan and Gaylord \cite{gaylord1} 
corroborated the existence of this state at 
$\bar \Gamma$ and present a discusson of  it (see Fig.~5 in Ref.~\cite{gaylord1}).

Figure~\ref{Ag001}(c) shows our calculated projected bulk band 
struture for Ag(111).
There we identify four SS labeled 
Es$_{2}$, Es$_{3}$, Es$_{8}$, and Es$_9$,  
and five RS labeled Er$_{1}$, $\ldots$, Er$_{5}$.
In general, from the figure we observe that the calculated projected
bulk bands structure are in agreement with that reported in Ref.~\cite{Ho1983}.
The characteristics for the different found SS and RS are
shown in Tables~\ref{tablaSS111} and \ref{tablaRS111}.

\subsubsection{Detailed discussion of the found SS and RS}

The SS Es$_{2}$ is located in the lower gap at --6.08 eV at $\bar \Gamma$.
The state shows a parabolic shape and its wavefunction 
is the hybridization of the $s,d_{3z^{2}-r^{2}}$ atomic orbitals.

Our found SS E$_{S3}$ was calculated at 2.28 eV at $\bar \Gamma$,
showing an important dispersion, its wavefunction has the symmetry
of the $s,p_{z}$ atomic orbitals.
From the experimental point of view a similar SS was reported by
Altmann {\it et al.} \cite{Altmann} and Reihl {\it et al.} \cite{frank2}
as an unoccupied state, while Kevan and Gaylord {\it et al.} \cite{gaylord1} 
reported the state as an occupied state.
Theoretically in Ref.~\cite{Ho1983}
the state was identified as an occupied state.

There are a couple of SS E$_{S8}$ and E$_{S9}$
located in the upper gap at $\bar K$.
E$_{S8}$ was found at 6.63 eV and its wavefunction has the 
symmetry of the $p_x,\ p_y$ atomic orbitals.
E$_{S9}$ was calculated at 6.97 eV, and its wavefunction has the 
symmetry of the $s,\ p_z$ atomic orbitals. 

The RS Er$_1$ was calculated at -6.42 eV at $\bar \Gamma$, and we 
found that the state has the symmetry of the $d_{3z^{2}-r^{2}}$
atomic orbitals.
Although there is no experimental evidence for this RS,
theoretically the state was predicted in Ref.~\cite{Ho1983}
(see Table~\ref{tablaRS111} for energy comparison).

In the energy bulk bands above the lower gap at $\bar \Gamma$,
we found the state Er$_2$ at -5.04 eV, the calculated wavefunction
for this RS was the hybridization of the $d_{xy,x^{2}-y^{2}}$
atomic orbitals.
The state was reported in Ref.~\cite{Ho1983}.

The state Er$_3$ was calculated at -6.23 eV at $\bar M$ point, 
and has the character of the $d$ orbitals.
Theoretically, this state was predicted in Ref.~\cite{Ho1983}.

The state Er$_4$  was calculated at --3.82 eV at $\bar M$ point,
and its wavefunction has the symmetry of the 
$ d_{xy,yz,zx,x^{2}-y^{2}} $ atomic orbitals.
Theoretically, the state was reported in Ref.~\cite{Ho1983}.

The state Er$_5$ was calculated in the lower border of the 
upper gap located in the $\bar M$ point. The state was located at
2.08 eV, and has the symmetry of the $s,/ p_x,\ p_y$ atomic
orbitals.

The states Es$_{3},$ Er$_{1},$ Er$_{2},$ Er$_{3}$ y Er$_{4}$ have been reported 
previously, and we found that our calculations agree with the 
reported values, with the exception of the energy values for E$_{S3}$.
The new states found in this work are 
 E$_{S2}$, and Er$_{5}$.

\begin{table}[!ht]
\caption{Surface states for the Ag(111). 
The first column is the found SS, the second column list the $k$-point 
where the SS was found, the third column shows the experimental 
energy reported for the related state, the fourth list the theoretical 
energy value reported in the literature for the SS, the next column 
shows our calculated energy value for the found SS, finally the last 
column shows the symmetry of atomic orbitals that form the SS, 
according to our calculation.}
\begin{tabular}{llllll}
\hline\hline
Es & $\vec{k}$ & E$_{\rm{exp}}$ & E$_{\rm{theo}}$ & E$_{\rm {our}}$ 
                                                  & SFO$_{\rm {our}}$ \\ \hline
Es$_{2}$ & $( 0,0) $ &  &  & -6.08 & $s,d_{3z^{2}-r^{2}}$ \\ 
         &           & 0.00\cite{Altmann} &  &  & \\
Es$_3$   & (0,0)     & 0.33\cite{frank2}   & -0.31\cite{gaylord1} & 2.28 & 
                                                  $s,p_z$ \\ 
         &           & -0.12\cite{gaylord1}  &   &  &  \\
Es$_{8}$ & $\left( \frac{4}{3}\sqrt{2},0\right) $ &  &  & $\ \ \ 6.63$ & $%
p_{x,y}$ \\ 
Es$_{9}$ & $\left( \frac{4}{3}\sqrt{2},0\right) $ &  &  & $\ \ \ 6.97$ & $%
s,p_{z}$ \\ 
\hline\hline
\end{tabular}
\label{tablaSS111}
\end{table}

\begin{table}[!ht]
\caption{Resonant states found for Ag(111) surface.
The first column shows the labeled resonance state, the 
second one shows the wavevector of the state in units of 
$[\frac{\pi}{a}]$, the next column shows the related energy in eV, 
finally the last column shows the wave symmetry found 
for the different states.}
\begin{tabular}{l l l l l}
\hline\hline
Er & $\vec{k}$ & E$_{\text{te\'{o}rico}}$ & E$_{\text{our}}$\textbf{\ } & SFO$%
_{\text{our}}$ \\ \hline
Er$_{1}$ & $( 0,0) $ & $-6.26$ \cite{Ho1983} & $-6.42$ & $d_{3z^{2}-r^{2}}$ \\ 
Er$_{2}$ & $( 0,0) $ & $-5.04$ \cite{Ho1983} & $-4.78$ & $d_{xy,x^{2}-y^{2}}$ \\ 
Er$_{3}$ & $( \sqrt{2},\frac{\sqrt{6}}{3}) $ & $-6.26$\cite{Ho1983} & $-6.01$ & 
                                                                   $d$ \\ 
Er$_{4}$ & $( \sqrt{2},\frac{\sqrt{6}}{3}) $ & $-3.82$\cite{Ho1983}
& $-3.36$ & $d_{xy,yz,zx,x^{2}-y^{2}}$ \\ 
Er$_{5}$ & $( \frac{4}{3}\sqrt{2},0) $ &  & $2.08$ & $s,\ p_x,\ p_y$ \\ 
%Er$_{6}$ & $( \frac{4}{3}\sqrt{2},0) $ &  & $-5.33$ & $%
%d_{yz,3z^{2}-r^{2}}$ \\ 
\hline\hline
\end{tabular}
\label{tablaRS111}
\end{table}

\end{document}